\documentclass[twoside,12pt]{article}
\usepackage{epsfig}

\def\Journal#1#2#3#4{{#1} {#2} (#4) #3 }

\def\NPA{{\em Nucl. Phys.} A}

\def\PLB{{\em Phys. Lett.} B}

\def\PRL{\em Phys. Rev. Lett.}

\def\PRC{{\em Phys. Rev.} C}

\newcommand{\be}{\begin{equation}}
\newcommand{\ee}{\end{equation}}
\newcommand{\bea}{\begin{eqnarray}}
\newcommand{\eea}{\end{eqnarray}}

\begin{document}

\title{ \vspace{1cm} Neutron Skin and Giant Resonances in Nuclei}
\author{Vadim Rodin
\\
Institute for Theoretical Physics, University of T\"ubingen\\
Auf der Morgenstelle 14, D-72076 T\"ubingen, Germany}

\maketitle

\begin{abstract} 
Some aspects, both experimental and theoretical, of extracting the neutron skin $\Delta R$ from properties of isovector giant resonances are discussed. Existing proposals are critically reviewed.
The method relying on the energy difference between the GTR and IAS is shown 
to lack sensitivity to $\Delta R$. 
A simple explanation of the linear relation between the symmetry energy and the neutron skin is also presented. 

\end{abstract}

\section{Introduction}

Accurate experimental data on the neutron skin in neutron rich nuclei would allow
to further constrain model 
parameters involved in the calculations of 
the nuclear symmetry energy~\cite{Horo01}. The latter plays a central role in a variety of
nuclear phenomena. The value $a_4 \approx $ 30 MeV of the nuclear symmetry energy 
$S(\rho_0) = a_4+ \frac{p_0}{\rho_0^2} (\rho-\rho_0)+\ldots$
at nuclear saturation density $\rho_0 \approx 0.17$ fm$^{-3}$ seems reasonably well established.
On the other hand, the 
 density dependence of the symmetry energy
can vary substantially with the  many-body approximations
employed. 

Several authors have pointed out~\cite{br00,furn02} a strong correlation
between the neutron skin, $\Delta R=\sqrt{\langle r^2\rangle_n}-\sqrt{\langle r^2\rangle_p}=R_n-R_p,$ 
and the symmetry energy of neutron matter near saturation density.
In the framework of a mean field approach Furnstahl~\cite{furn02} demonstrated that in
heavy nuclei there exists an almost linear empirical correlation
between theoretical predictions in terms of various mean field
approaches to $S(\rho)$  
(i.e., a bulk property)  and the neutron skin, $\Delta R$ (a
property of finite nuclei).

This observation has contributed to a renewed interest in an accurate
determination of the neutron skin in neutron rich nuclei. Besides, a precise value of the
neutron skin is required as an input in several processes of physical interest, e.g.
the analysis of energy shifts in deeply bound pionic atoms~\cite{Kolo}, and 
in the analysis of atomic parity violation experiments
(weak charge)~\cite{Poll99}. It is worth to stress that to experimentally determine the skin in heavy nuclei
is extremely challenging as $\Delta R$ is just about few percents of the nuclear 
radius.

The present contribution is partially based upon the results published previously in~\cite{diep03}.

\section{Relationship between the symmetry energy and $\Delta R$}\label{Relationship}

Brown \cite{br00} and Furnstahl~\cite{furn02} have pointed out that within the
framework of mean field models there exists an
almost linear empirical correlation between theoretical
predictions for both $a_4$ and its density dependence, $p_0, $    and
the neutron skin $\Delta R$ in heavy nuclei.
This observation suggests an intriguing
relationship between a bulk property of infinite nuclear matter
and a surface property of finite systems.
Here, following the analysis of~\cite{diep03}, this question is addressed from a point of view
of the Landau-Migdal approach. 

Let us consider a simple mean-field model
with the Hamiltonian consisting of the single-particle mean field
part $\hat H_0$ and the residual particle-hole interaction $\hat H_{p-h}$:
\begin{eqnarray}
&\hat H=\hat H_0+\hat H_{p-h},\ \ \ \ \hat H_{ph}=\sum\limits_{a>b}
(F'+G'\vec\sigma_a\vec\sigma_b)\vec\tau_a\vec\tau_b
\delta(\vec{r}_a-\vec{r}_b),\label{1.1}\\
&\hat H_0=\sum\limits_a (T_a+U(x_a)),\ \ \ \ U(x)=U_0(x)+U_1(x)+U_C(x),\\
&U_0(x)=U_0(r)+U_{so}(x);\ \ U_1(x)=\frac12 S_{\rm pot}(r)\tau^{(3)};\ \ 
U_C(x)=\frac12 U_C(r)(1-\tau^{(3)}).  
\end{eqnarray}
Here, $U_0(x)$ is the phenomenological isoscalar part of the mean field potential $U(x)\ (x=\{\vec r,\vec\sigma,\vec\tau\})$, 
$U_0(r)$ and $U_{so}(x)$ 
are the central and spin-orbit parts, respectively; $F'$ and $G'$ are the phenomenological Landau-Migdal parameters.
The isovector part $U_1(x)$ and the Coulomb mean field $U_C(x)$ are both calculated consistently in
the Hartree approximation, $S_{\rm pot}(r)$ is the symmetry potential 
($r$-dependent symmetry energy in finite nuclei). 

\begin{figure}[tb]
\begin{center}
\begin{minipage}[t]{8 cm}
\epsfig{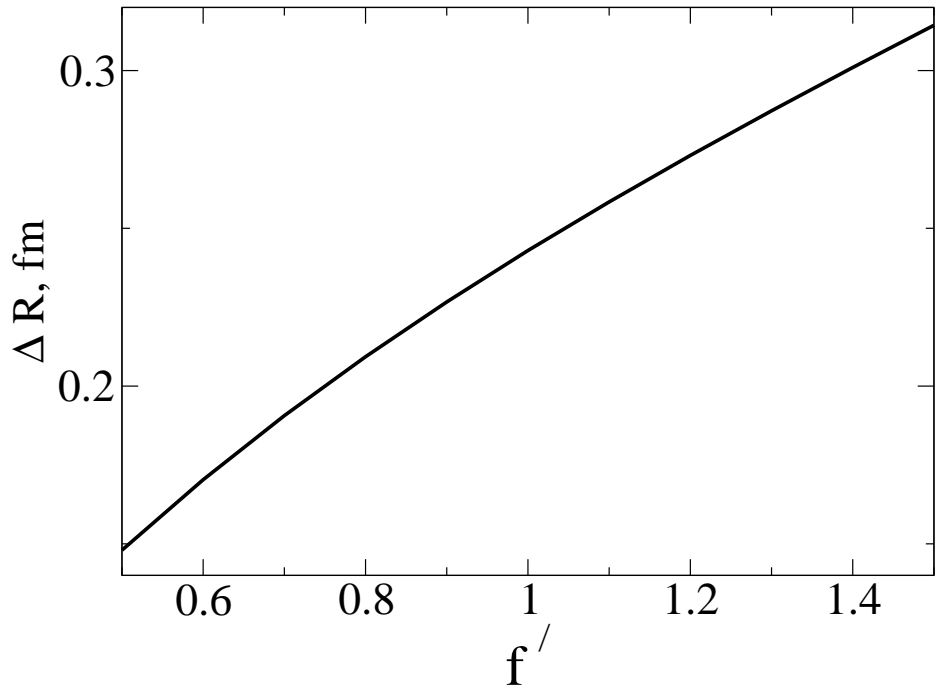}
\end{minipage}
\begin{minipage}[t]{16.5 cm}
\caption{Neutron skin in $^{208}$Pb versus the Landau-Migdal parameter $f'$.}
\label{DRvsF'}
\end{minipage}
\end{center}
\end{figure}

The model Hamiltonian $\hat H$ (\ref{1.1}) preserves the isospin
symmetry within the RPA if a selfconsistency relation between the symmetry potential 
and the Landau-Migdal parameter $F'$ is fulfilled:
\begin{eqnarray}
&S_{\mbox{pot}}(r)=2F'n^{(-)}(r), \label{1.5}
\end{eqnarray}
where $n^{(-)}(r)=n_{n}(r)-n_{p}(r)$ is the neutron excess density.
Thus, in this model the depth of the symmetry potential is
controlled by the Landau-Migdal parameter $F'$ (analogous
role plays the parameter $g_\rho^2$ in relativistic mean field models).
 $S_{\mbox{pot}}(r)$ is obtained from Eq.(\ref{1.5}) by an iterative procedure;
the resulting  dependence of $\Delta R$ on the dimensionless
parameter $f'=F'/ (300$ MeV\,fm$^3$)  shown in fig.~\ref{DRvsF'}
indeed illustrates that $\Delta R$ depends almost linearly on $f'$. 
Then with the use of the Migdal relation $a_4= \frac{\epsilon_F}{3}(1+2f')$~\cite{Mig83} relating 
the symmetry energy and $f'$, a similar, almost linear, correlation between 
$ a_4 $ and $ \Delta R $ is obtained.

To get more insight in the role of $f'$ we consider small variations $\delta F'$.
Neglecting  the variation of $n^{(-)}(r)$ with respect to $\delta F'$, the corresponding linear
variation of the symmetry potential is $\delta S_{pot}(r)=2\delta F'n^{(-)}(r)$. 
Then in the first order perturbation theory, such a variation of $S_{pot}$ causes the following 
variation of the ground-state wave function
$|\delta 0\rangle=\delta F'\sum\limits_{s} \frac{\langle s | \hat
N^{(-)} | 0\rangle}{E_0-E_s} |s\rangle$,
with $``s"$ labeling the eigenstates of the nuclear Hamiltonian and
a single-particle operator $\hat N^{(-)}$ defined as 
$\hat N^{(-)}=\sum\limits_a n^{(-)}(r_a)\tau^{(3)}_a$.
Consequently, the variation of the expectation value $\langle 0 | \hat V^{(-)} | 0 \rangle=N R^2_n-Z R^2_p$
of another single-particle operator $\hat V^{(-)}=\sum\limits_a r^2_a\tau^{(3)}_a$
can be written as
\begin{equation}
R_p \delta( \Delta R) =\delta F'\cdot\frac{2}{A}\sum\limits_{s}
\frac{{\rm Re}\langle 0 | \hat N^{(-)} | s\rangle \langle s | \hat
V^{(-)} | 0\rangle} {E_0-E_s}. \label{deltar2np}
\end{equation}
In practice the sum in Eq.~(\ref{deltar2np}) is exhausted mainly
by the isovector monopole resonance (IMR) which high excitation energy
(about 24 MeV in $^{208}$Pb) justifies the perturbative
consideration. Eq.~(\ref{deltar2np}) is able to
reproduce directly calculated $\delta (\Delta R)$ shown in
Fig.~\ref{DRvsF'} with the accuracy of about 10\%. As a result, a
simple microscopic interpretation of the linear correlation
between $\Delta R$ and Landau parameter $F'$ is obtained.

\section{Extracting neutron skin from properties of isovector giant resonances
} \label{Exp}

Parity violating electron scattering off nuclei is probably
the least model dependent approach
to probe the neutron distribution~\cite{Horopv}.  
The weak electron-nucleus potential is $\tilde{V}(r)= V(r)+\gamma_5A(r)$, 
where the axial potential $A(r)=\frac{G_F}{2^{3/2}} \rho_W(r)$.
The weak charge is mainly determined by neutrons $\rho_W(r)= ( 1-4\sin^2\theta_W)
\rho_p(r)- \rho_n(r)$, with $\sin^2\theta_W  \approx 0.23$. 
In a scattering experiment using polarized electrons one can
determine the cross section asymmetry~\cite{Horopv} which comes
from the interference between the $A$ and $V$ contributions. Using
the measured neutron form factor at small finite value of $Q^2$
and the existing information on the charge distribution one can
uniquely extract the neutron skin. Some slight model dependence
comes from the need to assume a certain radial dependence for the
neutron density,  to extract $R_n$ from a finite $Q^2$ form
factor. 

However, the best claimed accuracy of the experimental determination of neutron radii would be on the level of 1\%, 
that translates to relatively large uncertainty of 20-30\% in the neutron skin. On such accuracy level, some indirect 
experimental probes of $\Delta R$ still can be competitive.

A variety of experimental approaches have been employed 
to obtain indirect information on $\Delta R$. To some extent all
the analysis contain a certain model dependence, which in many cases is difficult to
estimate quantitatively. For choosing an indirect probe it is very important to address the question {\em how sensitive is the proposed physical quantity with respect to a variation of $\Delta R$ in a single nucleus}.
The higher is the sensitivity, the better is the choice of the correlation
for the indirect deducing $\Delta R$ from the measured values.

It is not intended here to give a comprehensive
review of the existing methods. 
In particular, the results from the analysis of the antiprotonic atoms, elastic proton and neutron 
scattering reactions, and the pygmy dipole resonance are completely left out.
Here, special emphasis will be put on proposals 
to provide accurate information on the neutron skin from properties of 
isovector giant resonances. 


\subsection{Spin-dipole Giant Resonance}
In \cite{kras99} it has been proposed to utilize the excitation probability of the spin-dipole resonance in charge
exchange reactions for determining the neutron skin. The method has been applied to obtain information
on the variation of the neutron skin in the Sn isotopes~\cite{kras99}.
For the relevant operator, $\sum_a \tau_a^{\pm} [\vec\sigma_a \otimes \vec r_a]_{JM},\ (J=0,1,2)$
the summed $\Delta L=1$ strength is 
  \begin{eqnarray}
& S^{(-)}-S^{(+)}=C(N R^2_n-Z R^2_p).
\label{1}
\end{eqnarray}
Here $S^{(-)}$ and $S^{(+)}$ are the spin-dipole total strengths
in $\beta^{(-)}$ and $\beta^{(+)}$ channels, respectively; $C$ is
the factor depending on the normalization of the spin-dipole operator
(in the definition of Ref.~[2] $C=1/4\pi$, we use here $C=1$).
Because $S^{(+)}$ could not be measured experimentally, the
model-dependent energy-weighted sum rule 
was invoked in the analysis of~\cite{kras99} to eliminate $S^{(+)}$.
However, the used analytical representation for the sum rule was oversimplified and
led in some cases, e.g. for $^{208}$Pb, to absurdly negative $S^{(+)}$. 
In~\cite{rod00} another way was proposed, namely, to use for the analysis the ratio $S^{(+)}/S^{(-)}$ calculated within the pn-RPA. The parameterization of the RPA calculation results for tin isotopes in the form 
\begin{eqnarray}
& S^{(+)}/S^{(-)} = 0.388 - 0.012(N-Z) \nonumber
\end{eqnarray}
was used later in ~\cite{kras02} to reanalyze the experimental data and led to a marked change in the extracted 
$\Delta R$'s.

Let us now assess the experimental accuracy for $S^{(-)}$ needed to determine
the neutron skin to a given accuracy. Putting $S^{(+)}=0$ (that seems to be a very good approximation 
for $^{208}$Pb) and one has
\begin{eqnarray}
& S^{(-)}=(N-Z) R^2_p+2N R_p\Delta R . \label{S-}
\end{eqnarray}
The ratio of the second term on the rhs to
the first one in case of  $^{208}$Pb is
$$ 2N\Delta R /((N-Z)  R_p)\approx 5.7 \Delta R / R_p. $$
Therefore, for $R_p=5.5$ fm and $\Delta R=0.2$ fm the second term is only 25\%
of the first one and one needs 5\% accuracy in $S^{(-)}$ to determine $\Delta R$ with
20\% accuracy. Because the SD strength is spread out and probably has a considerable
strength at low-energy,
the results for the $\Delta R$ can be only considered as
qualitative  with a relatively large uncertainty (up to 30-50\%).

\subsection{Isobaric analogue state}

The dominant contribution to the energy weighted sum rule (EWSR)
for Fermi excitations by the operator   $T^{(-)}=\sum_a \tau_a^{-}$
comes from the Coulomb mean field 
\be
(EWSR)_F= \int U_C(r)n^{(-)}(r)d^3r, \label{1.11}
\ee
The Coulomb mean field $U_C(r)$ resembles very much
that of the uniformly charged sphere, being inside a nucleus a quadratic
function: $U_C(r)=\displaystyle\frac{Ze^2}{2R_c}(3-(r/R_c)^2), \ r\le R_c$. It
turns out that
if one extends such a quadratic dependence also to the outer region  $r>R_c$
(instead of proportionality to $R_c/r$),
it gives numerically just a very small deviation in $(EWSR)_F$ (less than 0.5\%,
due to the fact, that both
the difference and its first derivative go to zero at $r=R_c$ and $n^{(-)}(r)$ is
exponentially decreasing for $r>R_c$). Using such an approximation, one gets:
\begin{eqnarray}
&&(EWSR)_F\approx (N-Z)\Delta_C\left(1-\frac{S^{(-)}}{3(N-Z)R^2_c}\right)
\label{1.22}
\end{eqnarray}
with $\Delta_C=\displaystyle\frac{3Ze^2}{2R_c}$, and $S^{(-)}$
given in Eq.(\ref{S-}).

Since the IAS exhausts almost 100\% of the NEWSR and EWSR, one may
hope to extract $S^{(-)}$ from the IAS energy. However, the term
depending on $S^{(-)}$ contributes only about 20\% to $(EWSR)_F$,
and as a result, the part of $S^{(-)}$ depending on $\Delta R$
contributes only about 4\% to $(EWSR)_F$ (in $^{208}$Pb). 
>From the experimental side, the IAS energy can be determined with
unprecendently high accuracy, better than
0.1\%. Also, from the experimentally known charge density distribution
 the Coulomb mean field $U_C(r)$ can be
calculated rather accurately, and hence one can determine
 the small difference between Eqs.(\ref{1.22}) and (\ref{1.11}).
But at the level of 1\% accuracy several theoretical effects
discarded in Eq.(\ref{1.11}) come into play
that makes such an accurate description of the IAS energy very difficult 
(the Nolen-Schiffer anomaly).

Also in~\cite{duf02} it was stated that the Coulomb displacement energies (CDE)
are sensitive to $\Delta R$. A gross estimate $\Delta R=0.80(5)(N-Z)/A$ fm
was obtained from a four-parameter fit of the experimental $R_p$ and observed mirror CDE's.
The authors claimed $127$ keV to be the rms error of the fit, but they assumed 
the nuclear wave functions calculated within the Nuclear Shell Model to be isospin pure.
Thus, the important effect of the Coulomb mixing of the IAS and the IMR was not taken into 
account, which is known to decrease the IAS energy by a few percents.
Therefore, the Nolen-Schiffer anomaly does not seem to have been resolved yet.

\subsection{Is the energy spacing between GTR and IAS a good candidate for determining the neutron skin in isotopic
chains?}

In a recent paper~\cite{Ring03} a proposal has been put forward to use
the isotopic dependence of the energy spacing, $\Delta E$, between the Gamow-Teller resonance (GTR)
and the IAS as a tool for determining the evolution
of the neutron skin in nuclei along an isotopic chain. Here, we would like
to present some physical arguments which question the physical relevance of this method. 

The authors of~\cite{Ring03} have used the fact that both functions, $\Delta R$ and $\Delta E$,
are monotonic functions (increasing and decreasing, respectively) of the neutron excess $(N-Z)$ to state
that ``isotopic dependence of the energy spacings between the GTR and IAS provides direct information on the evolution
of neutron skin-thickness along the Sn isotopic chain". Arguing in such a way one can find a correlation
between any two monotonic functions of a single physical parameter and plot them as a function of one another like
is done in Fig.~2 of~\cite{Ring03} \footnote{note that, to avoid confusion in comparing the measured and calculated dependences,
the authors should have plotted the experimental points in the upper panel as the function of the {\em measured} $\Delta R$
rather than {\em calculated} $\Delta R$ and should have added the horizontal error bars to them
reflecting the experimental uncertainty in $\Delta R$ shown in the lower panel.}.
However, it does not imply automatically a real physical correlation between the functions
which are determined also by many other model parameters which are kept fixed while performing calculations
(the calculations in~\cite{Ring03} have been performed within the relativistic mean field (RMF) and relativistic QRPA
(RQRPA) approaches).

Again, {\em the relevant question to be addressed is how sensitive is one physical quantity with respect to a variation of another in a single nucleus?}
In other words, one has to evaluate what variation of $\Delta E$ is produced by varying $\Delta R$ in a single nucleus.
Imaging an extreme situation (which is actually not far from reality) that $\Delta E$ were not
sensitive to $\Delta R$ at all, one would get by varying $\Delta R$ a family of different {\em calculated} dependences
(like shown in the upper panel of Fig.~2 of~\cite{Ring03}) which would give no clue about the {\em real} dependence
seen in nature.

Thus, it is quite important to understand the physical reasons which cause the energy splitting between the GTR and the IAS.
It is well-known that if the nuclear Hamiltonian possessed Wigner SU(4) symmetry then the GTR and the IAS would be degenerate, $\Delta E=0$.
In such a case any variation of $\Delta R$, not violating the symmetry, would not affect $\Delta E$ at all. However, it is also known
that the spin-isospin SU(4) symmetry is broken in nuclei. Hence, $\Delta E$ is determined by those terms in the nuclear 
Hamiltonian which violate the symmetry. Their qualitative and semi-quantitative estimates in terms of the energy weighted sum rules for the Gamow-Teller ($EWSR_{GT}$) and
the Fermi ($EWSR_{F}$) excitations have been already known for more than 20 years (see, e.g.,~\cite{GTR}).
The analysis of these authors as well as a quantitative analysis performed recently in~\cite{Rod03} 
has shown that there are three basic sources in the Hamiltonian which violate SU(4) symmetry
and contribute to the difference of the sum rules: spin-orbit mean field and both particle-particle and particle-hole residual charge-exchange interactions.
One sees that none of the sources explicitly refers to the symmetry potential, to which $\Delta R$ is especially sensitive.

An estimate of $\Delta E$ as $\Delta E=\displaystyle \frac{EWSR_{GT}-EWSR_{F}}{N-Z}$ 
can be calculated according to~\cite{Rod03} in the Sn isotopes. 
From the sources violating SU(4) symmetry, spin-orbit mean field represents the major one
and contributes about 5 MeV to the splitting. 
The contribution of the particle-hole interaction is negative and
about 1--2 MeV in the absolute value.  
The contribution of the particle-particle interaction is rather difficult to evaluate
(due to uncertainty in the strength of the spin-dependent particle-particle interaction)
but it seems to be of minor importance (very probably no more than 0.5 MeV, especially for large $(N-Z)$) 
and can safely be neglected.

Now let us turn to the discussion of the sensitivity of the contributions to the variation of $\Delta R$.
We could reproduce the corresponding analytical expressions from~\cite{Rod03} explicitly, 
but it is enough for our purpose just to mention
that the dominating contribution to $\Delta E$ from the spin-orbit mean field  is given by its expectation value
in the ground state and is determined basically only by the unfilled spin-orbit doublets. 
This expectation value is completely insensitive to the variation of $\Delta R$.


Within the Landau-Migdal approach described above, the particle-hole contribution 
\be 
\Delta E_{ph}=\displaystyle\frac{2(G'-F')}{N-Z}\int (n^{(-)}(r))^2 d^3r
\ee
is given by the product of the volume integral of the neutron excess density squared and the difference of the p-h strengths $G'$ and $F'$~\cite{Rod03}. In the $SU(4)$-symmetric limit one has $G'=F'$ and $\Delta E_{ph}=0$ explicitly. Still, in this limit one has a freedom to choose different $F'$ that produces a variation in $\Delta R$, similar to shown in fig 1. Therefore, as already mentioned, one can get no clue about the actual $\Delta R$ from $\Delta E=0$ in the $SU(4)$-symmetric limit.

In a realistic situation $G'\ne F'$  ($f'=1.0$ and $g'=0.8$ were taken in~\cite{Rod03}), but $\Delta E_{ph}$ depends
only on the difference $G'-F'$. One usually fixes $G'$ in order to reproduce the GTR energy in some nuclei (the authors of~\cite{Ring03} have followed this way, too) and possible information from $\Delta E$ about the absolute value of $F'$ is lost. Furthermore,
one can {\it a priori} think that a degree of violation of the SU(4) symmetry should be a sort of a fundamental property of the residual interaction.
Therefore, the difference $G'-F'$ should stay more stable in different models as compared to some possible variation
of $F'$ producing different $\Delta R$.

Considering value of $G'-F'$ fixed, one can employ a simple model varying only $\rho_n(r)$ to see 
how a change of $\Delta R$ affects $\Delta E_{ph}$ via variation of the neutron excess density $n^{(-)}(r)$.
A small variation of $\rho_n(r)$ can be approximately represented as $\delta\rho_n(r)=-\displaystyle\frac{\delta R_n}{R_n}
(3\rho_n(r)+R\displaystyle\frac{d\rho_n(r)}{dr})$,
where $\delta R_n$ is a change of the rms neutron radius $R_n$, $R$ is the nuclear radius (with $R_n^2\approx 0.6R^2$).
Assuming the proton and neutron densities be constant inside a nucleus, the final estimate is
$\displaystyle\frac{\delta\Delta E_{ph}}{\Delta E_{ph}}=-3\displaystyle\frac{\delta R_n}{R_n}(1+\frac{N + Z - 2\gamma N}{N-Z})$, where $\gamma=n^{(-)}(R)/n^{(-)}(0)$.

Thus, in Sn isotopes with the experimental charge radii about $R_p=$4.6 fm a rather significant variation of $\Delta R$ about 0.1 fm, that is of the order of magnitude of $\delta R$, would cause $\frac{\delta\Delta E_{ph}}{\Delta E_{ph}}=0.3$
and $\frac{\delta\Delta E_{ph}}{\Delta E_{ph}}=0.15$ for $^{112}$Sn and $^{132}$Sn, respectively ($\gamma=0.5$), that corresponds to the absolute change about 0.3 MeV in $\Delta E$, to be compared with the experimental uncertainties in $\Delta E$ of the same order. 
It is clear that to draw any conclusion about $\Delta R$ from the measured $\Delta E$ would be premature.
Even if the experimental errors in $\Delta E$ were exactly zero, the accuracy of the theoretical model itself
would be hardly believed to be of the necessary level. For instance, apart from the obvious uncertainties
in the isotopic dependence of the spin-orbit potential, the GTR does not exhaust 100\% of the corresponding
sum rules and the shell-structure effects such as configurational and isospin splitting of the GTR can have
some effect on the calculated GTR energy.   

It is also noteworthy that, in spite of the claimed self-consistency of the calculations, the slope of the calculated isotopic dependence of the IAS energy is about 3 times larger than
the experimental one (see inset in Fig.~1 of~\cite{Ring03}). Note, that the isospin self-consistent continuum-QRPA calculations of~\cite{Rod03} were able to nicely reproduce the slope
(while overall underestimated the IAS energy by about 0.5 MeV, the well-known Nolen-Schiffer anomaly).

To conclude, we believe that the suggested in~\cite{Ring03} method to deduce the neutron skin from
the energy spacing between GTR and IAS is rather questionable in its origin
and does not fairly provide ``direct information on the evolution of neutron skin-thickness".

\section{Some implications of $\Delta R$}

In several processes of physical interest
knowledge of  $\Delta R$ plays a crucial role and in fact a more
accurate value could lead to more stringent tests:

(i) The pion polarization operator \cite{Kolo} (the s-wave optical
potential) in a heavy nucleus 
$ \Pi(\omega,\rho_p,\rho_n)=-T^+(\omega)\rho -T^-(\omega) (
\rho_n-\rho_p) $ 
has mainly an isovector character ($T^+(m_\pi) \sim 0).$
Parameterizing the densities by Fermi shapes for the case of $^{208}$Pb the main
nuclear model dependence in the analysis comes  from the
uncertainty in  the value of $\Delta R$ multiplying $T^-.$

(ii) The parity violation in atoms is dominated by $Z-$boson
exchange between the electrons and the neutrons \cite{Poll99}. 
Taking the proton distribution as a reference there is a
small so-called neutron skin (ns) correction to the parity
non-conserving amplitude, $\delta E_{\mbox{pnc}}^{\mbox{ns}},$
for, say, a $6s_{1/2} \to 7s_{1/2}$ transition, which is related
to $\Delta R$ as 
(independent of the electronic structure) 
\be \frac {\delta E^{\mbox{ns}}_{\mbox{pnc}}}{
E_{\mbox{pnc}}}= -\frac{3}{7} (\alpha Z)^2 \frac {\Delta R}{R_p}.
\ee 
In $^{133}$Cs it amounts to a $\delta E/E \approx -(0.1-
0.4)\% $ depending on whether the non-relativistic or relativistic
estimates for $\Delta R $ are used \cite{Poll99}. The
corresponding uncertainty in the weak charge  $Q_W$ is
$-(0.2-0.8)\sigma.$

 (iii) The  pressure in a neutron star matter can be expressed as
in terms of symmetry energy and its density dependence 
\be P(\rho,x)= \rho^2 \frac{\partial E(\rho,x)} {\partial \rho} =
\rho^2 [ E'(\rho, 1/2) +S'(\rho) (1-2x)^2 +\ldots ]. \label{press}
\ee By using beta equilibrium in a neutron star, $\mu_e=
\mu_n-\mu_p= -\frac{\partial E(\rho,x)}{\partial x}, $ and the
result for the electron chemical potential, $\mu_e=3/4\hbar cx(3\pi^2\rho x)^{1/3}, $ one
finds the proton fraction at saturation density, $\rho_0,$ to be quite
small, $x_0 \sim 0.04.$
Hence, the pressure at saturation density  can be approximated as
\be P(\rho_0) = \rho_s(1-2x_0)(\rho_0S'(\rho_0)(1-2x_0)+ S(\rho_0)
x_0) \sim \rho^2_0 S'(\rho_0). \ee
At higher densities the proton fraction increases; this increase is more rapid in case of larger
$p_0$ \cite{Horo01}. While for the pressure at higher densities
contributions from other nuclear quantities like compressibility
will play a role in 
 it was argued that that there is a  correlation of the neutron star
 radius and the pressure 
which does not depend on the EoS at the highest densities.
 Numerically the correlation can be expressed in the form of a power law,
 $R_M \sim C(\rho,M) (\frac{P(\rho)}{\rm{MeV fm}^{-3}})^{0.25}$
 km, where $C(\rho=1.5 \rho_s,M=1.4 M_{solar})  \sim 7 . $ This shows
that a determination of a neutron star radius would provide some
constraint on the symmetry properties of nuclear matter.

\section{Conclusion}

In this contribution we discuss some aspects of extracting the neutron skin from properties of isovector giant resonances and critically review existing proposals.
The theoretical method relying on the energy difference between the GTR and IAS is shown 
to lack sensitivity to $\Delta R$. 
It is also shown that the phenomenological, almost linear, relationship
between the symmetry energy and the neutron skin in finite nuclei, observed
in mean field calculations, can be understood in terms the Landau-Migdal approach.

\subsection*{\it Acknowledgments}
The work is supported in part by the Deutsche Forschungsgemeinschaft (grant FA67/28-2) and 
by the EU ILIAS project (contract RII3-CT-2004-506222). The author would like to thank 
Profs. L.~Dieperink and M.~Urin for useful discussions.

\end{document}